\documentclass{iau} 
\usepackage{graphicx}
\usepackage{natbib}
\title[The multiplicity of massive stars: 2016] 
{The multiplicity of massive stars:\\a 2016 view}

\author[H. Sana]   
{Hugues Sana}

\affiliation{Institute of Astrophysics, KU Leuven, \\
Celestijnlaan 200D, 3001 Leuven, Belgium\\ 
email: {\tt hugues.sana@kuleuven.be} }

\pubyear{2017}
\volume{329}  
\jname{The lives and death-throes of massive stars}
\editors{A.C. Editor, B.D. Editor \& C.E. Editor, eds.}
\begin{document}
\def\aap{A\&A}
\def\aaps{A\&AS}
\def\aj{AJ}
\def\apj{ApJ}
\def\apjs{ApJS}
\def\apjl{ApJ}
\def\aplett{Astrophys. Lett.}
\def\araa{ARA\&A}
\def\mnras{MNRAS}
\def\nat{Nature}
\def\pasp{PASP}

\maketitle

\begin{abstract}
Massive stars like company. Here,   we provide a brief overview of progresses made over the last 5 years by a number of medium and large surveys.  These results  provide new insights on the observed and intrinsic multiplicity properties of main sequence massive stars and on the initial conditions for their future evolution. They also bring new interesting constraints on the outcome of the massive star formation process. 
\end{abstract}

\firstsection 
\section{Introduction}

 The presence of close companions introduces new physics that has the potential to affect all stages of  stellar evolution, from the pre-main sequence phase, to the type of end-of-life explosions and to the properties and orbital evolution of double-compact objects that may ultimately lead to detectable gravitational wave bursts. In this short paper, we  highlight major observational advances that have occurred in the last five years, with no ambition to provide a complete historical review. More information on works published prior to 2012 can be found in e.g., \cite{MHG09}, \cite{SaE11} and references therein. As in our previous report, we will focus on the initial multiplicity conditions, i.e. that of main-sequence massive stars.\\

\begin{figure}[t!]
\begin{center}
 \includegraphics[width=5.3in]{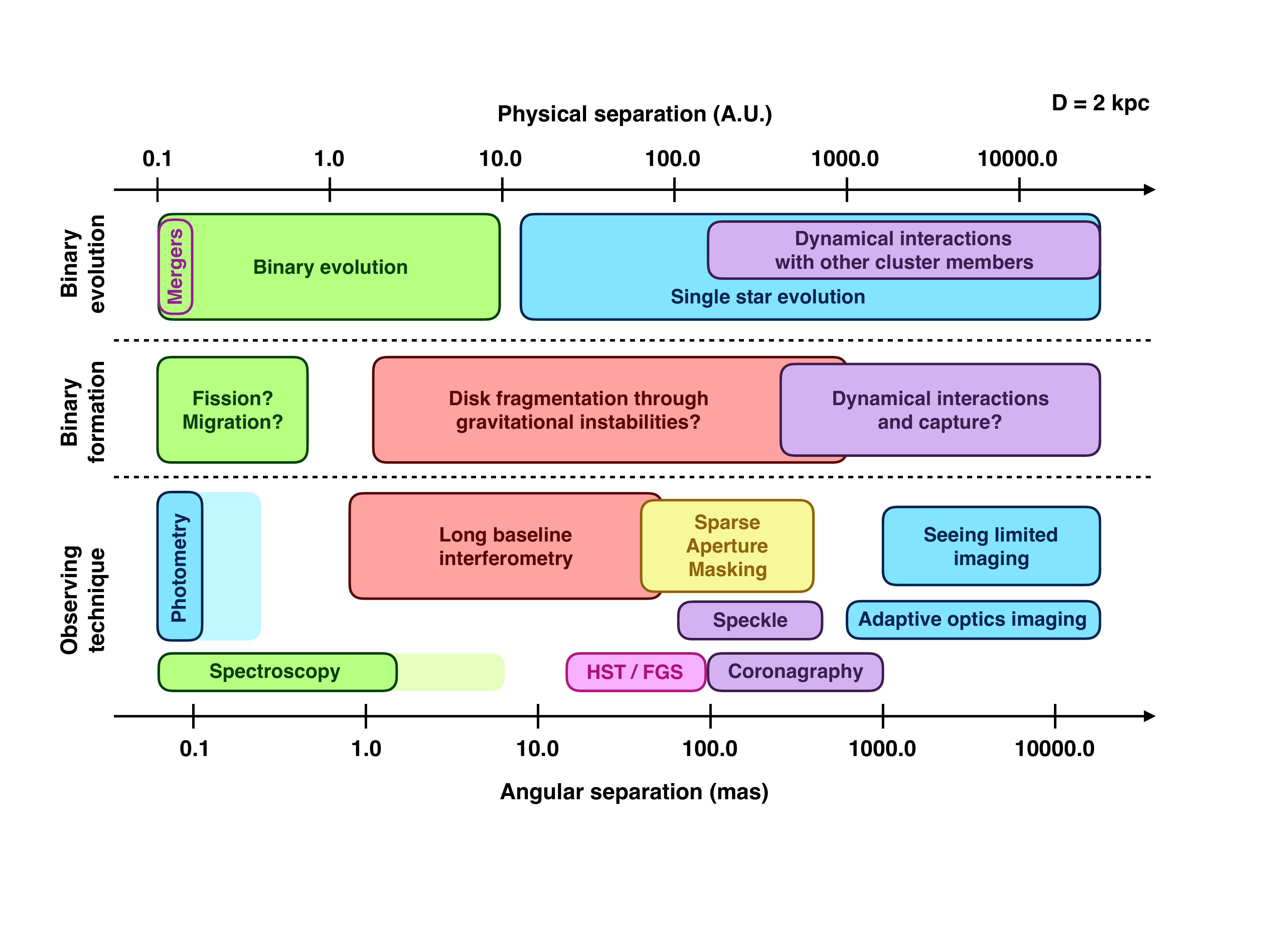} 
 \vspace*{-1.0 cm}
 \caption{Sketch of the typical physical and angular separations covered by O-type binaries. A distance of 2~kpc is adopted, as representative of most galactic multiplicity surveys. Domains of sensitivity for various instrumental techniques and relevant parameter space of  formation and evolution scenarios are indicated. Lightly shaded areas indicate regions where spectroscopy and photometry are in principle sensitive but where the detection likelihood drops dramatically. 
}
   \label{f:param}
\end{center}
\end{figure}

As  described by, e.g., \cite{MGH98}, \cite{SaE11} and  \cite{Moe},  four or five orders of magnitudes in separation need to be investigated to cover the entire parameter space relevant for the formation and evolution of massive stars  (Fig.~\ref{f:param}). This can only be achieved by a combination of instrumental techniques that  all have their own regime of optimal sensitivities (masses, brightnesses, separation, flux contrasts, mass-ratios, ...).  Recent advances is long-baseline interferometry -- mostly the magnitude limit and efficiency of the observations (see Sect. \ref{s:interfero}) -- have allowed us to bridge the spectroscopic and imaging regimes (the so-called interferometric gap, \cite{MGH98}). However, this is currently only possible for distances of up to $\sim$4~kpc as, farther away, most massive stars become too dim for current interferometric instrumentation. \\

In this short overview, we will focus on the spectroscopic (Sect.\ref{s:spectro}) and interferometric (Sect.~\ref{s:interfero}) regimes, with emphasis on surveys that have not been discussed elsewhere in these proceedings (see the contributions of Barb\'a, Sim\'on-D\'iaz and Vink). We will follow the terminology adopted in \cite{SLBL14}:\\
\begin{enumerate}
\item[-] {\bf The number of observed targets} or {\bf sample size} ($N$),
\item[-] {\bf The number of multiple systems } ($N_\mathrm{m}$):   number of observed central objects with at least one companion,
\item[-] {\bf  The fraction of multiple systems} or {\bf multiplicity fraction}: $f_\mathrm{m} = N_\mathrm{m}/N$, 
\item[-] {\bf The number of companions} ($N_\mathrm{c}$):  total number of companions observed around a given sample of $N$ central objects,
\item[-] {\bf The fraction of companions} or {\bf companion fraction} ($f_\mathrm{c} =N_\mathrm{c}/N$): average number of companions per central object.
\end{enumerate}

Depending on the context, these quantities can be  restricted to sub-categories, such as spectroscopic binaries systems (SBs) (Sect.~\ref{s:spectro}) or to specific separation ranges (Sect.~\ref{s:interfero}).

\section{Spectroscopic surveys}\label{s:spectro}

\begin{table}
  \begin{center}
  \caption{Overview of recent spectroscopic surveys and their observed ($f_\mathrm{SB}^\mathrm{obs}$) and bias-corrected ($f_\mathrm{SB}^\mathrm{corrected}$) binary fraction. The sample size ($N$), number of detected binaries ($N_\mathrm{SB}$) and number of binaries with an orbital solution ($N_\mathrm{SBO}$) are also given.  Error bars are computed following binomial statistics.}
  \label{t:spectro}
 {\scriptsize
  \begin{tabular}{|l|c|c|c|c|c|c|l|}\hline 
{\bf Region/Survey} & {\bf SpT} &{\bf $N$} & {\bf $N_\mathrm{SB}$} & {\bf $N_\mathrm{SBO}$} &{\bf $f_\mathrm{SB}^\mathrm{obs}$} &  {\bf $f_\mathrm{SB}^\mathrm{corrected}$}  & {\bf Reference} \\ 
    \hline
\multicolumn{8}{|c|}{Milky Way: O-type stars$^a$}\\
    \hline
BESO  Survey     & O       &  243  &     166   &  --  & $0.68\pm0.03$ & -- & \cite{CHN12}  \\
Young Clusters    & O       &  71   &         40  & 34 & $0.56\pm0.06$ & $0.69\pm0.09$ & \cite{SdMdK12} \\
GOSSS               & O       & 194 & 97 (low)  &   -- & $0.50\pm0.03$ &   -- & \cite{SMAM14} \\
                            & O       &  & 117 (high)  &   -- & $0.60\pm0.03$ &   -- & \\
 IACOB$^b$               & O       &   141 & 66 & -- & $0.47\pm0.04$ &   -- & Sim\'on-D\'iaz (these proceedings)  \\
Clusters/Assoc    & O       & 161 &    68 (low)  & 68 & $0.42\pm0.04$ &   -- & \cite{ACNG15} \\
                           & O       &      &      96 (high)  & 68 & $0.60\pm0.04$ &   -- &  \\
 OWN                  & O        & 205 & 102         & 85 & $0.50\pm0.03$     &     -- & Barb\'a (these proceedings) \\
 Cyg OB2            & O        & 45  &  23          & 23   &  $0.51\pm0.07$  & --$^c$ & \cite{KK14}\\
                             & {\it OB}     & {\it 128}  & {\it  48}          & {\it 48}   &  {\it 0.38~$\pm$~0.04}  & {\it 0.55$^c$} &\\
  \hline
\multicolumn{8}{|c|}{Milky Way: B-type stars}\\
    \hline
Cyg OB2                & B0-2   & 83  &  25          & 25   &  $0.30\pm0.05$  & --$^c$ & \cite{KK14} \\
    BESO Survey     & B0-3   &  226  &     105   &  --  &        $0.46\pm0.03$ & -- & \cite{CHN12} \\
                            & B4-9   &  353  &     67   &  --  &        $0.19\pm0.02$ & -- & \\
 \hline
\multicolumn{8}{|c|}{Large Magellanic Cloud}\\
    \hline
 30Dor/VFTS       & O        & 360 & 124         &  --   & $0.35\pm0.03$    &  $0.51\pm0.03$ & \cite{SdKdM13} \\
 30Dor/TMBM      & O        &        &                &  79  &   --      &  $\approx 0.58$ & \cite{AST17} \\
 30Dor/VFTS       &  B0-3  & 408 & 102 & -- & $0.25\pm0.02$ & $0.58\pm0.11$ & \cite{DDS15} \\
\hline
  \end{tabular}
  }
 \end{center}
\vspace{1mm}
 \scriptsize{
 {\it Notes:} $^a$The O-type samples in the Milky have strong overlap, so that the quoted measurements should not be viewed as being independent. 
$^b$Numbers  only concerns O stars with more than 3 observational epochs. The IACOB survey covers another $\sim40$ Northern O stars, but with fewer epochs so far.
$^c$\cite{KK14} do not provide separate completeness correction for the  O- and B-type sub-samples in Cyg OB2.   }
\end{table}

\subsection{Galactic surveys}

 In the last few years, two kinds of Galactic surveys have brought our understanding of massive star multiplicity to a new level:  
 \begin{enumerate} 
\item[(i)]  surveys focusing on specific young clusters (\cite{SdMdK12}) or OB association (Cyg OB2; \cite{KiK12,KK14}). These surveys have acquired a large number of observational epochs per target, allowing them to obtain orbital solutions for most of the detected spectroscopic binaries in their sample. They can further be considered as volume limited surveys in their own targeted regions. 

\item[(ii)] Galaxy-wide surveys with various spectral resolving power ($R=\lambda/\delta \lambda$) and number of epochs, and which are mostly magnitude limited, e.g.:
\item[\vspace*{3mm}- ] {\bf The Galactic O-Star Spectroscopic Survey} (GOSSS; \cite{SMAW11,SMAM14}, \cite{MASA16}): a (mostly) single-epoch survey at $R\approx 2500$ , which main objective is to bring to a firm ground the spectral classification of galactic O stars down to $B=13$;
 \item[-] {\bf The BESO Survey} of Galactic O and B stars (\cite{CHN12}): a $R=50\,000$ multi-epoch campaign that has allowed to search for radial-velocity (RV) variations in the observed sample;
 \item[-] {\bf The IACOB survey} (\cite{SDNMA15} and references therein, see also these proceedings):  multi-epoch high-resolution spectroscopic survey of Northern Galactic O- and B-type stars active since 2008; 
  \item[-] {\bf The OWN survey} (\cite{BGA14}, see also Barb\'a in these proceedings): a high-resolution spectroscopic monitoring of Southern O and WN stars with enough multi-epoch observations to measure the orbital properties of most detected binaries.
\end{enumerate}

\subsection{The Tarantula region}

Simultaneously to these groundbreaking observational campaigns in the Milky Way, the {\bf VLT-Flames Tarantula Survey} (VFTS, \cite{ETHB11}; see Vink et al., these proceedings) has systematically investigated about 800 massive OB and WR stars  in the 30 Dor region in the Large Magellanic Cloud. Using 6-epochs spectroscopic at moderate resolving power $R\sim 7000$, the VFTS  -- and its sequel program, the {\bf Tarantula Massive Binary Monitoring} (TMBM, \cite{AST17}) --- has provided direct observational constraints on the massive stars multiplicity properties in a metallicity environment that is reminiscent of that of more distant galaxies at $z\sim 1-2$.\\

Recent results of the VFTS are described in Vink et al.\ (these proceedings). Here we provide  more insight into the TMBM project.  TMBM  was designed to obtain multi-epoch spectroscopy of O-type binaries in the 30 Dor region with sufficient time sampling to measure the orbital properties of objects with orbital periods of up to about one year. TMBM targeted 60\%\ of every O-type binary candidates detected in the VFTS (see \cite{SdKdM13}) -- i.e., 93 out of 152 -- and obtained orbital solution for 78 of them. While specific objects are still being analysed,  results so far have allowed to identify two very interesting binaries, for which (quasi)-chemically homogeneous evolution ((q)CHE) seems to be required  to explain their observed properties. In both cases, (q)CHE may  affect the binary evolutionary path, by preventing the merging (VFTS 352) or Roche lobe overflow (R145):\\

\begin{enumerate}
\item[-] {\bf VFTS 352} is an overcontact 29+29~M$_\odot$ short period system  ($P_\mathrm{orb}=1.12$~d) that shows evidence of enhanced internal mixing possibly putting both components on a (q)CHE track (\cite{vfts352}). Ongoing abundance measurements may help to validate the existence of such an alternative scenario (\cite{MdM16}) to produce massive black holes with properties similar to that  at the origin of the recent gravitational wave detection (\cite{AAA16_gw}, \cite{MdM16}). \\
\item[-] {\bf RMC 145 (aka R145)} is  a WN6h + O3.5If$^*$/WN7 system with a high eccentricity ($e=0.78$) and rather low inclination ($i=39\pm6^\mathrm{o}$). Once suggested to contain a 300~M$_\odot$ star,  \cite{SRS17} showed that the current masses were most likely of the order of 80~M$_\odot$. Best estimates for initial masses yield values of 105 and 90~M$_\odot$ for the primary and secondary component respectively. Comparison with evolution tracks and the system high eccentricity suggest (q)CHE, allowing the system to have avoided episodes of mass-transfer so far.
\end{enumerate}

\subsection{The spectroscopic binary fraction}\label{s:fbin}

The previously described spectroscopic surveys of O and B stars, which have  samples of tens to hundreds of stars, yield  detected spectroscopic binary fractions  ($f_\mathrm{SB}^\mathrm{obs}$) that range from $f_\mathrm{SB}^\mathrm{obs}\sim0.25$ for the B-type stars in the 30 Dor region to about $f_\mathrm{SB}^\mathrm{obs}\sim0.7$ for O stars  in the Milky Way (see Table~\ref{t:spectro}). Much of this range can be explained by the nature and quality of the data (signal-to-noise ratio, number of epochs, ...) that directly impact the RV accuracy achieved by these surveys, hence the probability to detect significant Doppler shifts. Overall, all galactic O-star surveys yield a detected spectroscopic binary fraction $f_\mathrm{SB}^\mathrm{obs}> 0.5$, setting a firm lower limit on the intrinsic multiplicity fraction. The most optimistic estimates even reach $f_\mathrm{SB}^\mathrm{obs}\sim0.6$ to $0.7$ before any bias correction. \\

The overall binary detection probability  of these  surveys are difficult to estimate because it requires the knowledge of the parent orbital distributions of the targeted sample -- which are of course ill-constrained --  and a model of the sensitivity of each survey. Recent estimates yield an overall detection probability of spectroscopic binaries with $P_\mathrm{orb}$ up to $\approx 10$~years  of the order of 0.65 to 0.8 for O-type stars (\cite{SdMdK12, SdKdM13, KK14}). Corresponding intrinsic binary fractions are then in the range of  $f_\mathrm{SB}^\mathrm{corrected} \approx 0.5$ to  $0.7$. It remains to be seen whether the differences observed between the Milky Way and 30 Dor surveys are indeed genuine, possibly resulting from  environmental or from stellar (binary) evolution effects.\\

For B-type stars, $f_\mathrm{SB}^\mathrm{obs}$ rather lays in the range of 0.30 to $0.46$. For these stars, the detection rate is expected to be lower owing to fainter objects and smaller Doppler shifts, on average, resulting from smaller masses involved. Detailed considerations of the observational biases are needed to  better quantify the intrinsic B-type star multiplicity fraction.  Interestingly, measurements in the 30 Dor region show that, once biases are taken into account,  the O- and the B-type binary fractions are compatible within errors, with both mass ranges yielding  $f_\mathrm{SB}^\mathrm{corrected}\approx 0.58$ (admittedly errors remain quite large for B stars, see Table~\ref{t:spectro}).

\begin{figure}[t!]
\begin{center}
 \includegraphics[width=2.6in]{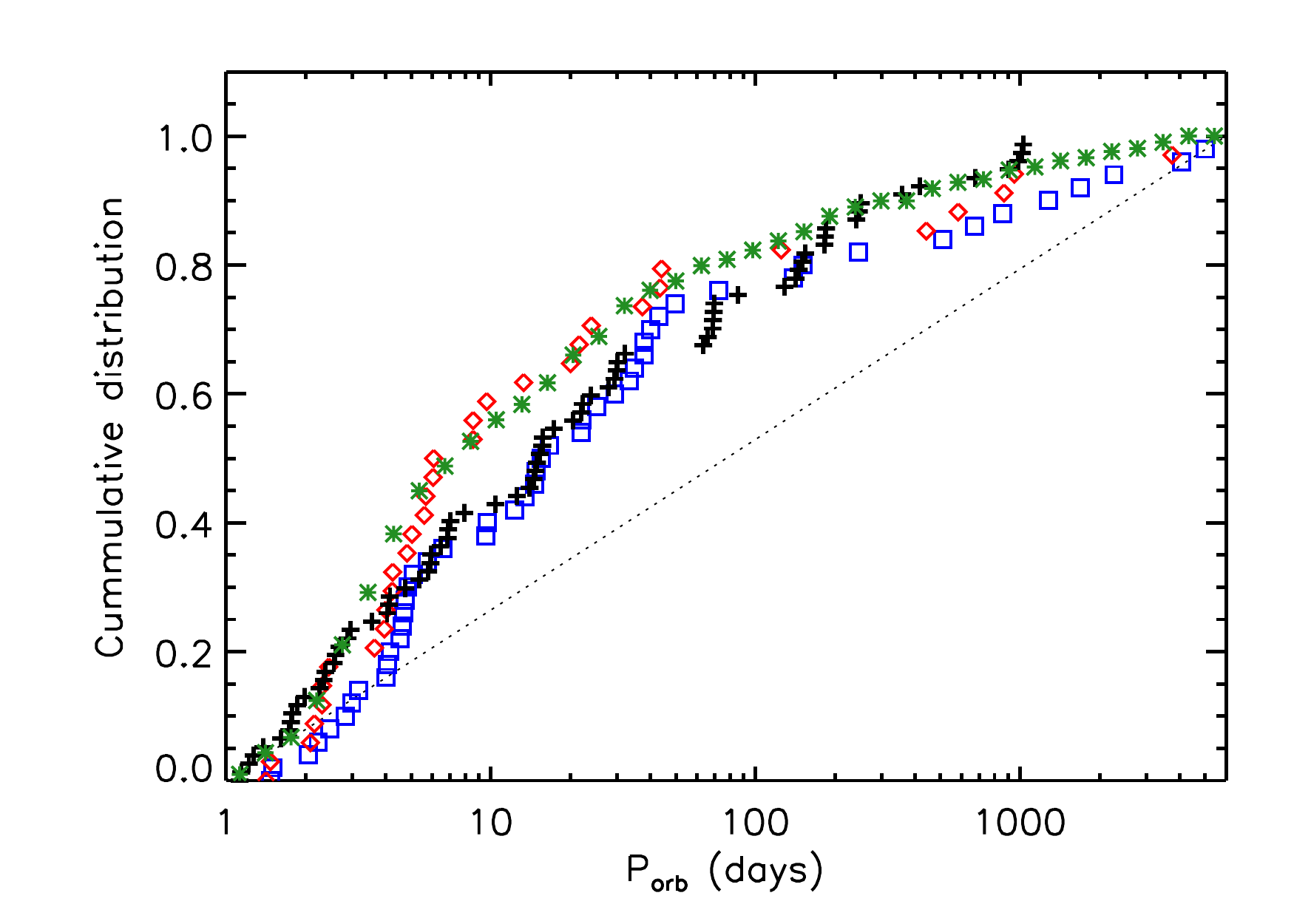}
 \includegraphics[width=2.6in]{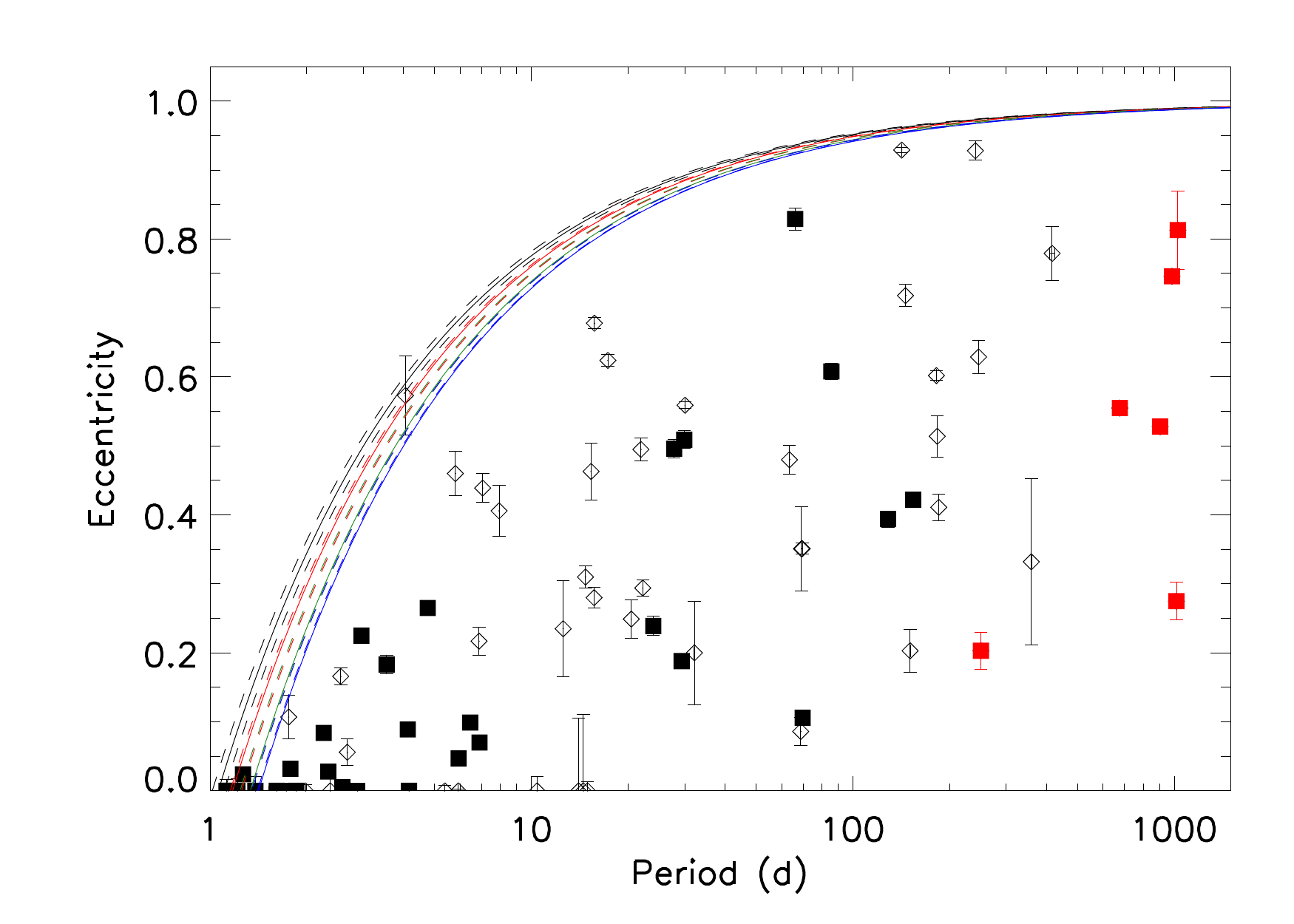}
 \caption{Left. Observed cumulative distributions of the orbital periods from different surveys: the young clusters (\cite{SdMdK12}, red diamonds), Cyg OB2 (\cite{KK14}, blue squares), 30 Dor/TMBM (\cite{AST17}, black crosses) and OWN (Barb\'a, these proceedings; private communication of the period histogram). The dotted line indicates a flat distribution in $\log_{10} P_\mathrm{orb}$.  Right. Eccentricity vs.\ period diagram for the TMBM O-star sample. Filled and open symbols indicates SB2 and SB1 systems, respectively. The red hexagons are for TMBM low-priority targets, i.e.\ stars showing $\Delta$RV$<20$~km\,s$^{-1}$ in VFTS. Dotted lines show the locus at which the two stars would touch at periastron for a representative range of stellar-masses and mass-ratios. Eccentricity measurement for the SB1 system at $P\sim4$~d and $e\sim0.6$ is biased and the data point should be neglected (see \cite{AST17}).}
   \label{f:period}
\end{center}
\end{figure}

\subsection{The orbital periods}\label{s:periods}

The initial orbital periods -- and their overall distribution -- are one of the most important quantities to properly predict the pre-supernova evolution of binary systems. Indeed, $P_\mathrm{orb}$ is the prime parameter that defines whether a system will interact through mass exchange. It sets the timing of the interaction and, to a large degree, its nature. Currently, only a few surveys have gathered enough observational epochs on a sufficiently large sample of binaries to construct observed period distributions. Constraints on the period distribution can in principle be obtained from modelling the distribution of the RV-variations (e.g., \cite{SdKdM13,DDS15}), but this requires assumptions on, e.g., the range of orbital periods (see  discussion in \cite{AST17}). \\

So far, three published surveys bring direct constraints on the orbital period distributions. These are displayed in Fig.~\ref{f:period} and show a large degree of similarities, despite the rather different environments in terms of stellar density and metallicity. Further more, preliminary results of the OWN survey (Barb\'a, these proceedings)  confirm and improve the statistics for Galactic massive stars. Future work will allow to investigate whether the small differences seen in Fig.~\ref{f:period} are statistically significant or whether they result from different observational biases, sample sizes and selection effects. In the former case, these may hold clues to the formation process and early dynamical evolution of massive stars as suggested by, e.g., \cite{KK14}.\\

The systematic and long-term nature of the  OWN and VFTS/TMBM surveys have allowed to detect and characterise an unprecedented number of long period systems. Thanks to these efforts, the period-eccentricity diagram, once showing a dearth of long-period low-eccentricity systems (see e.g., \cite{SdMdK12}), is now  well populated (Fig.~\ref{f:period} for TMBM, see also Barb\'a, these proceedings). The presence of these systems argues against dynamical capture and suggests that most of these spectroscopic binaries have been created during the formation process. 
\vspace*{-2mm}

\begin{figure}[t!]
\begin{center}
 \includegraphics[width=2.6in]{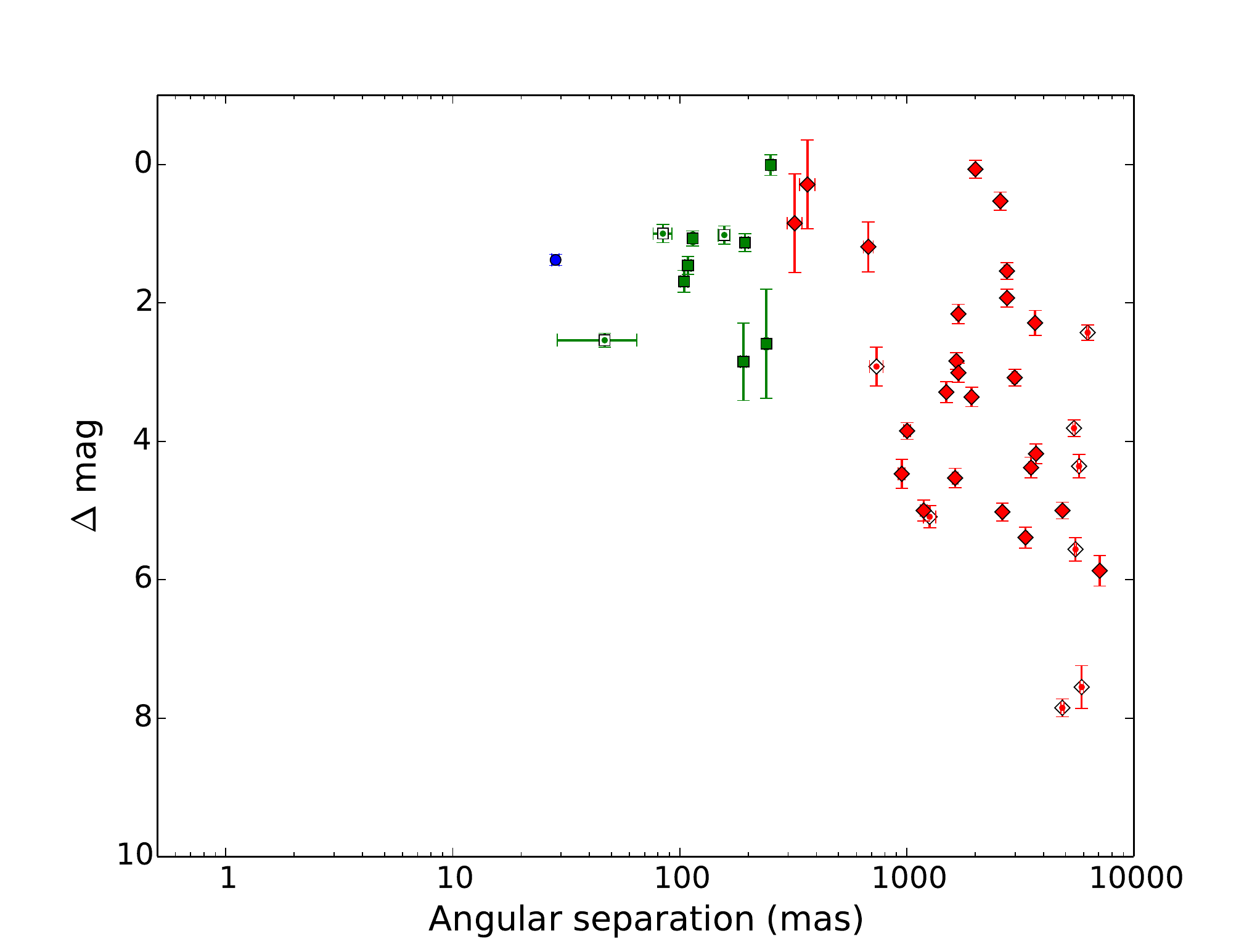}
  \includegraphics[width=2.6in]{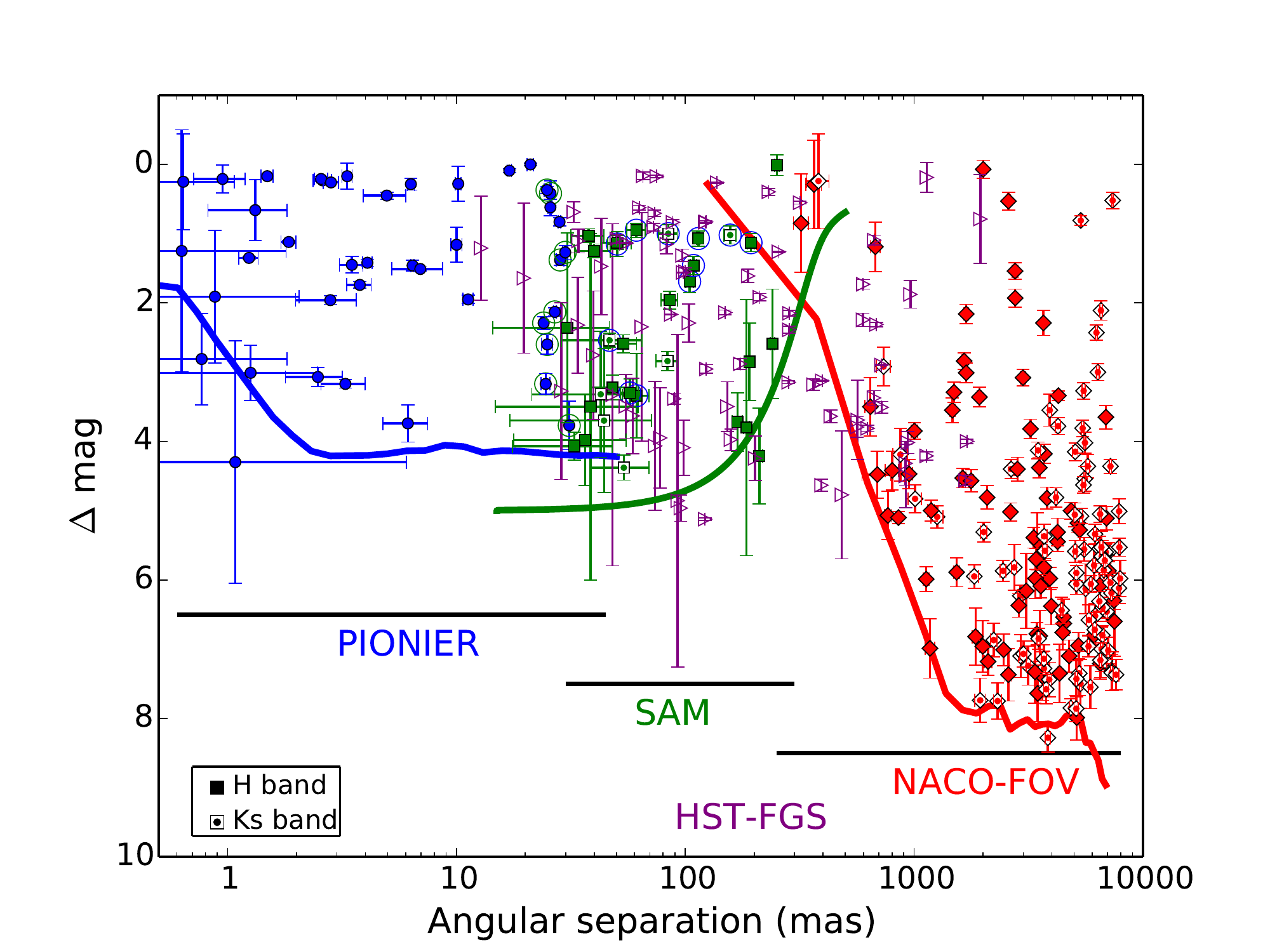}
 \caption{Left. Companions to the SMASH+ targets resolved before 2011. Right. Companions detected by the SMASH+ Survey (PIONIER: blue, NACO-SAM: green, NACO-FOV: red) in the H- (filled symbols) and Ks-bands (dotted-open symbols), and with the HST-FGS (purple) in the V-band (open symbols).  Plain lines show the median sensitivity limits of the different legs of the SMASH+ survey. Companions detected  both by the SMASH+ and the HST-FGS surveys appear twice in the figure. Based on data from \cite{SLBL14} and \cite{ACNG15}.}
   \label{f:smash}
\end{center}
\end{figure}

\section{High-angular resolution surveys}\label{s:interfero}

Two important Galactic surveys have allowed to systematically bridge the interferometric gap that was separating the spectroscopic and imaging domains (\cite{MGH98}, see also Fig.~\ref{f:param}). 
Both surveys show that the interferometric gap contains a significant number of companions and confirm previous suspicions that {\it the end product of massive star formation is a multiple system}.\\

\begin{enumerate}
\item[-] {\bf The Southern MAssive Star at High angular resolution} (SMASH+, \cite{SLBL14})  is an ESO  Large Program that targeted O-type stars in the Southern sky ($\delta < 0^\mathrm{o}$). SMASH+ combined long baseline interferometry (VLTI/PIONIER; 117 objects), aperture masking (VLT/NACO-SAM; 162 objects) and adaptive optics (VLT/NACO-FOV; 162 objects) to search for companions with separations in the range of 1 to 8000 mas. 264 companions were detected, of which  almost 200 were previously unresolved. After excluding runaway stars that show no resolved companions, 55\% of the 96 targets observed with the full suite of instruments revealed a companion in the range of 1 to  200~mas. Combined with existing spectroscopy (see Sect.~\ref{s:spectro}),  the SMASH+ results  yield a multiplicity fraction at angular separation $\rho < 8$" of  $f_\mathrm{m} = 0.90 \pm 0.03$ and an averaged number of companions of $f_\mathrm{c} = 2.1 \pm 0.2$. Interestingly,  all dwarfs stars in the SMASH+ sample  have a bound companion within $\sim 100$~AU. \\

\item[-] {\bf The HST fine guidance sensor  survey} (HST-FGS, \cite{ACNG15}) observed 224 O- and B-type stars in both hemispheres. The survey allowed to resolved 58 multiple systems (incl.\  43 new detections) with typical separation in the range of $\approx 20$ to 1000~mas. Focusing on the 214 stars in clusters and associations, the HST-FGS resolved a companion for 31\%\ of them. The authors performed their own literature review (see also Table~\ref{t:spectro}) and obtained $0.5 < f_\mathrm{m} < 0.7$ and $0.7<f_\mathrm{c} = 1.7$.  The HST-FGS survey nicely covers the sensitivity gap between the NACO-SAM and FOV techniques in the SMASH+ survey and shows that companions are also found at these separations.
\end{enumerate}

\begin{figure}[t!]
\begin{center}
  \includegraphics[width=2.6in]{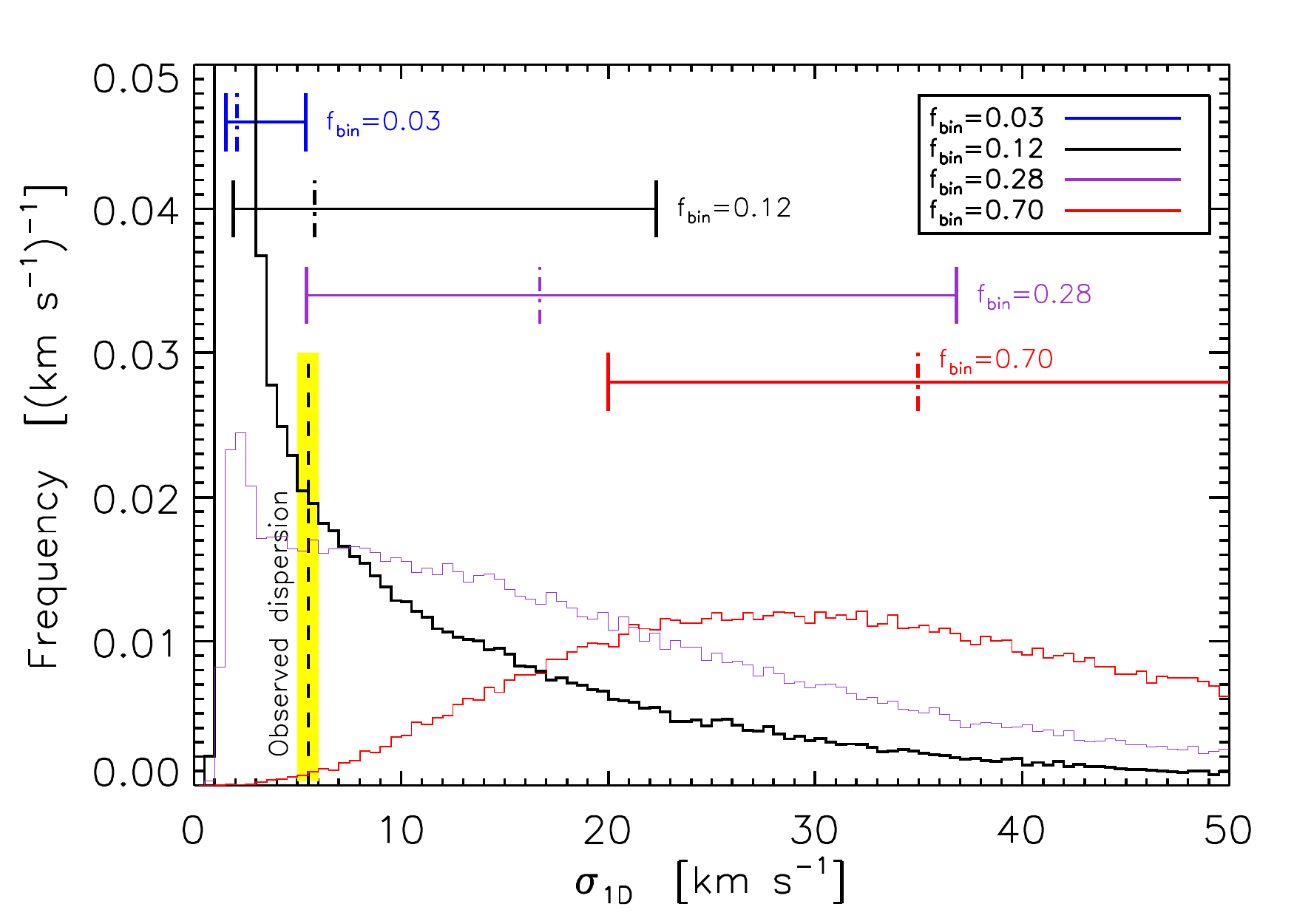}
  \includegraphics[width=2.6in]{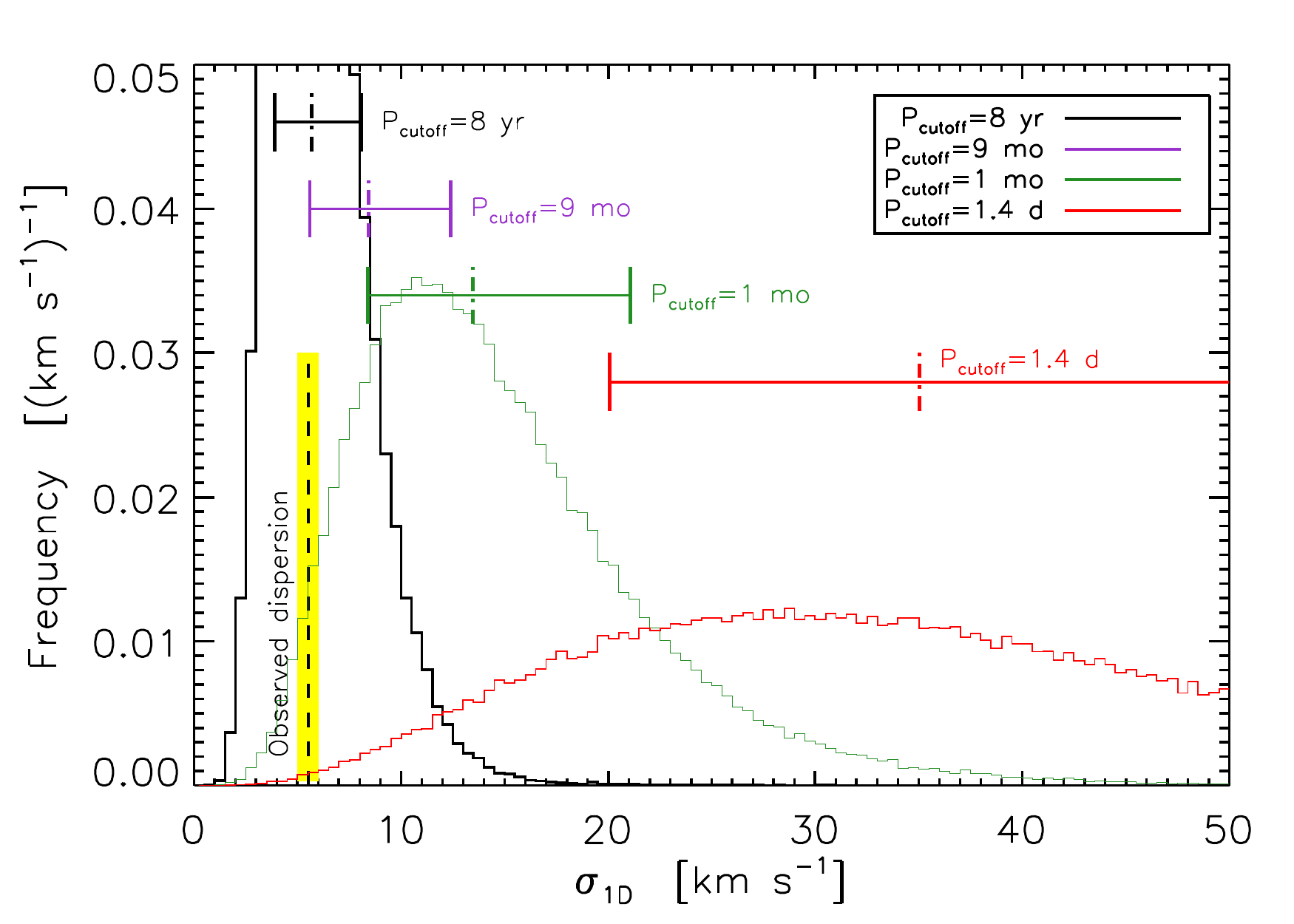}
 \caption{Simulated distribution of RV-dispersion ($\sigma_\mathrm{1D}$) for different multiplicity  properties of the parent populations.  Left. The binary fraction is varied. Right. The minimum orbital period ($P_\mathrm{cutoff}$) in the period distribution is varied. The shaded area (yellow) shows the RV-dispersion observed in the Ram\'irez-Tannus et al.\ sample. Figures adapted from \cite{SRTdK17}. }
   \label{f:m17}
\end{center}
\end{figure}

\section{News from M17: clues to the origin of massive binaries?}\label{s:m17}

Current constraints on the multiplicity of massive stars show that about half of them belong to  binary systems with  orbital periods of less than one month. This corresponds to separations below $\sim 1$~AU. The existence of such tight binaries is currently not predicted by massive star formation theories. In this context, the recent Xshooter spectroscopic campaign towards a dozen  young massive stars and mYSOs in the 1~Myr-old M17 region by Ram\'irez-Tannus et al.\  (\cite{Maria2017}, see also these proceedings) is worth mentioning. Most objects were observed only once. While no double-lined profiles were detected, the sample shows an intriguingly low RV-dispersion ($\sigma_\mathrm{1D}$), of the order of 5.5~km\,s$^{-1}$ (Fig.~\ref{f:m17}) only while values of 20 to 40~km\,s$^{-1}$ would have been expected for the typical multiplicity properties. \\

Subsequent Monte-Carlo analysis confirms that such a low $\sigma_\mathrm{1D}$ is indeed incompatible with the multiplicity properties of well characterised massive star populations such as those presented in Table~\ref{t:spectro}. Our results point either to a very low binary fraction ($f_\mathrm{SB} \sim 0.12$ only) or to a lack of short period binaries ($P_\mathrm{cutoff}> 9$~months). The first scenario would require to invoke that star formation has had a different outcome in M17 than in other OB-star regions in the Milky Way. The latter scenario may suggest that binaries first form at larger separations and then quickly harden to meet the observational constraints derived from slightly older populations (Table~\ref{t:spectro}). Such migration scenario may be driven by interaction with other companions in a multiple systems, or with remnant of the accretion disk. The  cut-off period may then be related to physical length-scales representative of the bloated pre-main-sequence stellar radii or of their accretion disks. The full discussion is to be found in \citet{Maria2017} and \cite{SRTdK17}.

\section{Final word}\label{ccl}
The results obtained in the last 5 years have definitely established the importance of massive star multiplicity. New large surveys have resulted in unprecedentedly high multiplicity and companion fraction, with values as large as $f_\mathrm{m}>0.9$ and $f_\mathrm{c}>2$ being obtained before bias-correction when considering the entire range of separations.  Of particular interest is the relative similarities of the multiplicity properties measured so far for different samples, despite different environments, sample ages and even metallicities. The one exception is the results from the M17 region (see Sect.~\ref{s:m17}).  More work is definitely required  to further explore the possible effects of age and metallicity on the multiplicity properties. As a final note, we insist on the critical need for observers to characterise the various observational biases as carefully as possible, including selection effects and sensitivity limits. While time consuming to establish, these quantities are indeed critical to retrieve the intrinsic multiplicity properties from the observations. A posteriori attempts to combine different samples may turn more complicated without this specific information at hand (see e.g., \cite{Moe}).

\section*{Ackowledgments}
The author is thankful to Rodolfo Barb\'a and  Sim\'on-D\'iaz for sharing OWN and IACOB results prior to publication (OWN results were used in the preliminary comparison shown in Fig.~\ref{f:period}) and to colleagues and collaborators for many stimulating discussions.


\begin{thebibliography}{22}
\expandafter\ifx\csname natexlab\endcsname\relax\def\natexlab#1{#1}\fi

\bibitem[{{Abbott} {et~al.}(2016){Abbott}, {Abbott}, {Abbott}, {Abernathy},
  {Acernese}, {Ackley}, {Adams}, {Adams}, {Addesso}, {Adhikari}, \&
  et~al.}]{AAA16_gw}
{Abbott}, B.~P., {Abbott}, R., {Abbott}, T.~D., {et~al.} 2016, Physical Review
  Letters, 116, 061102

\bibitem[{{Aldoretta} {et~al.}(2015){Aldoretta}, {Caballero-Nieves}, {Gies},
  {Nelan}, {Wallace}, {Hartkopf}, {Henry}, {Jao}, {Ma{\'{\i}}z Apell{\'a}niz},
  {Mason}, {Moffat}, {Norris}, {Richardson}, \& {Williams}}]{ACNG15}
{Aldoretta}, E.~J., {Caballero-Nieves}, S.~M., {Gies}, D.~R., {et~al.} 2015,
  \aj, 149, 26


\bibitem[Almeida et al.(2015)]{vfts352} Almeida, L.~A., Sana, H., de Mink, S.~E., et al.\ 2015, \apj, 812, 102 

\bibitem[{{Almeida} {et~al.}(2017){Almeida}, {Sana}, {Taylor}, {Barb{\'a}},
  {Bonanos}, {Crowther}, {Damineli}, {de Koter}, {de Mink}, {Evans}, {Gieles},
  {Grin}, {H{\'e}nault-Brunet}, {Langer}, {Lennon}, {Lockwood}, {Ma{\'{\i}}z
  Apell{\'a}niz}, {Moffat}, {Neijssel}, {Norman}, {Ram{\'{\i}}rez-Agudelo},
  {Richardson}, {Schootemeijer}, {Shenar}, {Soszy{\'n}ski}, {Tramper}, \&
  {Vink}}]{AST17}
{Almeida}, L.~A., {Sana}, H., {Taylor}, W., {et~al.} 2017, \aap, 598, A84 

\bibitem[{{Barb{\'a}} {et~al.}(2014){Barb{\'a}}, {Gamen}, {Arias}, {Morrell},
  {Walborn}, {Ma{\'{\i}}z Apell{\'a}niz}, {Sota}, \& {Alfaro}}]{BGA14}
{Barb{\'a}}, R., {Gamen}, R., {Arias}, J.~I., {et~al.} 2014, in Revista
  Mexicana de Astronomia y Astrofisica Conference Series, Vol.~44, 148--148

\bibitem[{{Chini} {et~al.}(2012){Chini}, {Hoffmeister}, {Nasseri}, {Stahl}, \&
  {Zinnecker}}]{CHN12}
{Chini}, R., {Hoffmeister}, V.~H., {Nasseri}, A., {Stahl}, O., \& {Zinnecker},
  H. 2012, \mnras, 424, 1925

\bibitem[{{Dunstall} {et~al.}(2015){Dunstall}, {Dufton}, {Sana}, {Evans},
  {Howarth}, {Sim{\'o}n-D{\'{\i}}az}, {de Mink}, {Langer}, {Ma{\'{\i}}z
  Apell{\'a}niz}, \& {Taylor}}]{DDS15}
{Dunstall}, P.~R., {Dufton}, P.~L., {Sana}, H., {et~al.} 2015, \aap, 580, A93

\bibitem[{{Evans} {et~al.}(2011){Evans}, {Taylor}, {H{\'e}nault-Brunet},
  {Sana}, {de Koter}, {Sim{\'o}n-D{\'{\i}}az}, {Carraro}, {Bagnoli}, {Bastian},
  {Bestenlehner}, {Bonanos}, {Bressert}, {Brott}, {Campbell}, {Cantiello},
  {Clark}, {Costa}, {Crowther}, {de Mink}, {Doran}, {Dufton}, {Dunstall},
  {Friedrich}, {Garcia}, {Gieles}, {Gr{\"a}fener}, {Herrero}, {Howarth},
  {Izzard}, {Langer}, {Lennon}, {Ma{\'{\i}}z Apell{\'a}niz}, {Markova},
  {Najarro}, {Puls}, {Ramirez}, {Sab{\'{\i}}n-Sanjuli{\'a}n}, {Smartt},
  {Stroud}, {van Loon}, {Vink}, \& {Walborn}}]{ETHB11}
{Evans}, C.~J., {Taylor}, W.~D., {H{\'e}nault-Brunet}, V., {et~al.} 2011, \aap,
  530, A108

\bibitem[{{Kiminki} \& {Kobulnicky}(2012)}]{KiK12}
{Kiminki}, D.~C. \& {Kobulnicky}, H.~A. 2012, \apj, 751, 4

\bibitem[{{Kobulnicky} {et~al.}(2014){Kobulnicky}, {Kiminki}, {Lundquist},
  {Burke}, {Chapman}, {Keller}, {Lester}, {Rolen}, {Topel}, {Bhattacharjee},
  {Smullen}, {Vargas {\'A}lvarez}, {Runnoe}, {Dale}, \& {Brotherton}}]{KK14}
{Kobulnicky}, H.~A., {Kiminki}, D.~C., {Lundquist}, M.~J., {et~al.} 2014,
  \apjs, 213, 34

\bibitem[{{Ma{\'{\i}}z Apell{\'a}niz} {et~al.}(2016){Ma{\'{\i}}z
  Apell{\'a}niz}, {Sota}, {Arias}, {Barb{\'a}}, {Walborn},
  {Sim{\'o}n-D{\'{\i}}az}, {Negueruela}, {Marco}, {Le{\~a}o}, {Herrero},
  {Gamen}, \& {Alfaro}}]{MASA16}
{Ma{\'{\i}}z Apell{\'a}niz}, J., {Sota}, A., {Arias}, J.~I., {et~al.} 2016,
  \apjs, 224, 4

\bibitem[{{Mandel} \& {de Mink}(2016)}]{MdM16}
{Mandel}, I. \& {de Mink}, S.~E. 2016, \mnras, 458, 2634

\bibitem[{{Mason} {et~al.}(1998){Mason}, {Gies}, {Hartkopf}, {Bagnuolo}, {ten
  Brummelaar}, \& {McAlister}}]{MGH98}
{Mason}, B.~D., {Gies}, D.~R., {Hartkopf}, W.~I., {et~al.} 1998, \aj, 115, 821

\bibitem[{{Mason} {et~al.}(2009){Mason}, {Hartkopf}, {Gies}, {Henry}, \&
  {Helsel}}]{MHG09}
{Mason}, B.~D., {Hartkopf}, W.~I., {Gies}, D.~R., {Henry}, T.~J., \& {Helsel},
  J.~W. 2009, \aj, 137, 3358

\bibitem[{{Moe} \& {Di Stefano}(2016)}]{Moe}
{Moe}, M. \& {Di Stefano}, R. 2016, arXiv: 1606.05347

\bibitem[Ram{\'{\i}}rez-Tannus et al.(2017)]{Maria2017} Ram{\'{\i}}rez-Tannus, M.~C., Kaper, L., de Koter, A., et al.\ 2017, \aap, in press (arXiv:1704.08216)

\bibitem[{{Sana} \& {Evans}(2011)}]{SaE11}
{Sana}, H. \& {Evans}, C.~J. 2011, in IAU Symposium, ed. {C.~Neiner, G.~Wade,
  G.~Meynet, \& G.~Peters}, Vol. 272, 474

\bibitem[{{Sana} {et~al.}(2012){Sana}, {de Mink}, {de Koter}, {Langer},
  {Evans}, {Gieles}, {Gosset}, {Izzard}, {Le Bouquin}, \&
  {Schneider}}]{SdMdK12}
{Sana}, H., {de Mink}, S.~E., {de Koter}, A., {et~al.} 2012, Science, 337, 444

\bibitem[{{Sana} {et~al.}(2013){Sana}, {de Koter}, {de Mink}, {Dunstall},
  {Evans}, {H{\'e}nault-Brunet}, {Ma{\'{\i}}z Apell{\'a}niz},
  {Ram{\'{\i}}rez-Agudelo}, {Taylor}, {Walborn}, {Clark}, {Crowther},
  {Herrero}, {Gieles}, {Langer}, {Lennon}, \& {Vink}}]{SdKdM13}
{Sana}, H., {de Koter}, A., {de Mink}, S.~E., {et~al.} 2013, \aap, 550, A107

\bibitem[{{Sana} {et~al.}(2014){Sana}, {Le Bouquin}, {Lacour}, {Berger},
  {Duvert}, {Gauchet}, {Norris}, {Olofsson}, {Pickel}, {Zins}, {Absil}, {de
  Koter}, {Kratter}, {Schnurr}, \& {Zinnecker}}]{SLBL14}
{Sana}, H., {Le Bouquin}, J.-B., {Lacour}, S., {et~al.} 2014, \apjs, 215, 15

\bibitem[Sana et al.(2017)]{SRTdK17} Sana, H., Ram{\'{\i}}rez-Tannus, M.~C., de Koter, A., et al.\ 2017, \aap, 599, L9 

\bibitem[{{Shenar} {et~al.}(2017){Shenar}, {Richardson}, {Sablowski},
  {Hainich}, {Sana}, {Moffat}, {Todt}, {Hamann}, {Oskinova}, {Sander},
  {Tramper}, {Langer}, {Bonanos}, {de Mink}, {Gr{\"a}fener}, {Crowther},
  {Vink}, {Almeida}, {de Koter}, {Barb{\'a}}, {Herrero}, \& {Ulaczyk}}]{SRS17}
{Shenar}, T., {Richardson}, N.~D., {Sablowski}, D.~P., {et~al.} 2017, \aap,
  598, A85

\bibitem[Sim{\'o}n-D{\'{\i}}az et al.(2015)]{SDNMA15} Sim{\'o}n-D{\'{\i}}az, S., Negueruela, I., Ma{\'{\i}}z Apell{\'a}niz, J., et al.\ 2015, Highlights of Spanish Astrophysics VIII, 576 

\bibitem[{{Sota} {et~al.}(2011){Sota}, {Ma{\'{\i}}z Apell{\'a}niz}, {Walborn},
  {Alfaro}, {Barb{\'a}}, {Morrell}, {Gamen}, \& {Arias}}]{SMAW11}
{Sota}, A., {Ma{\'{\i}}z Apell{\'a}niz}, J., {Walborn}, N.~R., {et~al.} 2011,
  \apjs, 193, 24

\bibitem[{{Sota} {et~al.}(2014){Sota}, {Ma{\'{\i}}z Apell{\'a}niz}, {Morrell},
  {Barb{\'a}}, {Walborn}, {Gamen}, {Arias}, \& {Alfaro}}]{SMAM14}
{Sota}, A., {Ma{\'{\i}}z Apell{\'a}niz}, J., {Morrell}, N.~I., {et~al.} 2014,
  \apjs, 211, 10

\end{thebibliography}

\end{document}